# The Governance Inversion Hypothesis: Why More AI Regulation May Produce Less Organisational Control

**Victor Frimpong**[1]

Abstract

This paper introduces the Governance Inversion Hypothesis (GIH) to explain a growing paradox in artificial intelligence (AI) governance: under conditions of increasing regulatory expansion and technological complexity, organisations may become more formally governed while simultaneously experiencing a decline in operational control over AI systems. Existing AI governance frameworks generally assume that stronger regulation improves accountability, oversight, and organisational control. This paper challenges that assumption by arguing that governance formalisation contributes to the erosion of control in AI-intensive environments. Drawing on institutional theory, organisational governance research, accountability scholarship, and emerging AI governance literature, the paper develops a conceptual framework that explains how regulatory expansion weakens operational authority through four interconnected mechanisms: *authority fragmentation, symbolic governance expansion, externalisation of control*, and *authority paralysis*. As governance systems become increasingly layered and procedurally dense, organisations may struggle to maintain coherent authority, technical visibility, escalation capability, and meaningful intervention power over opaque and externally mediated AI infrastructures. The paper extends institutional decoupling theory by introducing governance inversion as a structural condition in which governance expansion actively undermines operational coherence rather than strengthening it. It concludes that the central risk in AI governance may not be the absence of governance structures but the emergence of institutions that appear increasingly governed while progressively losing the capacity to govern effectively.

## INTRODUCTION

The governance of artificial intelligence (AI) is increasingly recognised as essential in highly regulated sectors, including banking, healthcare, insurance, and public administration. Regulatory bodies and governments are raising expectations around accountability, transparency, fairness, explainability, and human oversight. Recent initiatives such as the EU AI Act (European Union, 2024) and the NIST AI Risk Management Framework (NIST, 2023) reflect a growing policy and academic consensus that stronger governance requirements improve organisational control over AI systems.

[1] DBA, Lecturer and Member of the Research Advisory Board, SBS Swiss Business School, Kloten-Zurich, Switzerland; email: v.frimpong@research.sbs.edu

This assumption appears increasingly unstable. As AI systems expand into vital functions such as credit assessment, fraud detection, diagnostics, cybersecurity, recruitment, and public services, organisations increasingly rely on opaque models, third-party infrastructures, cloud ecosystems, and external AI services within technologically intensive risk societies characterised by escalating institutional complexity and uncertainty (Beck, 1992). Despite expanding governance expectations, many organisations continue to experience fragmented authority, unclear accountability, and limited operational oversight over these systems (Frimpong & Botchey, 2026; Papagiannidis et al., 2025), raising a structural question: *can expanding governance frameworks unintentionally weaken organisational control in AI-intensive environments?*

This paper introduces the Governance Inversion Hypothesis (GIH) to address that paradox. It argues that as AI complexity and regulatory demands increase, broader governance frameworks can inadvertently reduce effective control over AI systems, through four interconnected mechanisms: authority fragmentation, symbolic governance expansion, externalisation of control, and authority paralysis.

Current AI governance research places heavy emphasis on ethical principles, regulatory frameworks, and accountability, often assuming organisations already have the capacity to implement governance effectively (Morley et al., 2020; Wieringa, 2020). This paper instead treats AI governance as a structural problem of organisational control — one that depends not only on formal structures but on authority, technical understanding, operational unity, and the ability to intervene in complex AI systems. Methodologically, the paper adopts a theory-building conceptual approach grounded in institutional theory, organisational governance literature, and emerging AI governance research, arguing that in AI-intensive environments, greater governance formalisation does not always translate into greater governance capacity.

## LITERATURE REVIEW

### AI Governance and the Assumption of Organisational Control

The rapid development of artificial intelligence (AI) governance has produced a growing body of literature on accountability, transparency, fairness, explainability, auditability, and human oversight (Floridi & Cowls, 2019; Mittelstadt, 2019; OECD, 2022). Recent frameworks such as the NIST AI Risk Management Framework, ISO/IEC 42001, and the EU AI Act focus on structured governance systems, formal accountability, risk assessment, and organisational oversight (NIST, 2023; ISO, 2023a, 2023b).

This literature has advanced discussions of governance principles (Floridi & Cowls, 2019), operational tools (Morley et al., 2020), algorithmic accountability (Raji et al., 2020), socio-technical oversight (Rahwan, 2018), and governance frameworks (Papagiannidis et al., 2025), shifting AI governance from theoretical ethics toward practical application. However, much of this work assumes organisations already possess the institutional capacity to implement governance effectively, treating accountability as a design issue rather than a capacity issue. In practice, AI systems often operate through opaque processes, adaptive learning, and third-party or vendor-controlled infrastructures that limit internal visibility and control (Burrell, 2016; Koshiyama et al., 2024), so governance structures can grow without a corresponding increase in oversight capability.

Existing research on AI governance therefore overlooks a critical issue: the balance between governance frameworks and the actual authority organisations hold over AI systems. Much of the literature focuses on ideal governance principles rather than on how organisational structures shape regulation and control in practice (Prem, 2023; Wieringa, 2020) — implicitly treating governance structures as equivalent to governance power, when establishing committees, ethics frameworks, and compliance mechanisms does not guarantee the capacity to control increasingly autonomous AI systems.

The study re-examines the relationship between governance expansion and organisational control under conditions of AI complexity, external dependency, and fragmented institutional authority.

**Accountability, Organisational Design, and Structural Fragmentation**

Accountability is central to discussions of modern AI governance. Current frameworks emphasise clear responsibility, auditability, explainability, and effective human oversight (Floridi & Cowls, 2019; OECD, 2022), typically achieved through mechanisms such as transparency requirements, documentation standards, algorithmic audits, risk assessments, ethics committees, and reporting procedures (Raji et al., 2020; Shneiderman, 2020).

Accountability in AI systems, however, is as much a structural and organisational challenge as a procedural one. Classical accountability theory assumes responsibility can be clearly defined and allocated within structured hierarchies (Bovens, 2007), but this breaks down in AI-driven organisations where decision-making is spread across technical teams, compliance functions, external vendors, cloud services, data providers, legal units, and operational managers. Research on algorithmic accountability confirms these challenges: opaque, adaptive AI processes make it difficult to trace causality or identify who holds actual authority over model behaviour, decision outputs, system changes, or governance actions (Burrell, 2016; Wieringa, 2020).

This fragmentation is evident across highly regulated sectors like banking, healthcare, insurance, and public administration where multiple oversight functions (risk, compliance, legal, ethics, technology, internal audit, external regulators) coexist. While intended to strengthen oversight, these structures can weaken operational authority by spreading responsibility without clear decision-making ownership. Institutional theory helps explain why: organisations often adopt formal governance structures to appear legitimate under external pressure, even when those structures are not effectively integrated into operations (Meyer & Rowan, 1977), a dynamic intensified in AI governance, where organisations face strong pressure to visibly demonstrate responsible practice to regulators, investors, clients, and the public.

AI governance further overlaps with existing organisational functions — compliance, legal, risk management, technology operations — creating confusion about decision-making authority even as it improves coordination. This problem deepens with external dependency: reliance on cloud platforms, proprietary models, vendor algorithms, and third-party APIs limits an organisation's visibility into system behaviour, even though the organisation remains formally accountable for outcomes, producing a structural asymmetry between institutional accountability and operational authority.

Governance structures thus expand even as operational authority becomes more fragmented and harder to manage — a shift that motivates re-examining whether expanded governance frameworks inadvertently reduce effective control in AI-intensive systems.

**Regulatory Complexity, Proceduralisation, and Control Capacity**

Contemporary frameworks prioritise documentation, auditability, transparency, explainability, risk classification, impact assessments, and human oversight (European Union, 2024; NIST, 2023). As AI systems are increasingly used in high-risk sectors, organisations must demonstrate compliance through governance committees, reporting systems, ethics reviews, algorithmic audits, vendor assessments, and formal oversight processes.

This proceduralisation reflects a broader trend in which organisations rely on formal procedures, documentation, and audits to reduce uncertainty and enhance legitimacy (Power, 1997; Black, 2008). Yet research on institutional complexity indicates that added governance layers do not necessarily improve operational effectiveness: in environments with high institutional density and technological uncertainty, they can instead create coordination challenges, fragmented authority, and slower decision-making (Crozier, 1964; Perrow, 1984).

The tension is sharper in AI-intensive environments, where adaptive, rapidly evolving systems built on dynamic optimisation and distributed infrastructure routinely outpace traditional governance processes (Rahwan, 2018; Koshiyama et al., 2024). Rather than redesigning authority relationships, institutions typically respond by expanding procedural governance itself, producing overlapping compliance, legal, ethics, cybersecurity, audit, procurement, and operational functions.

Excessive procedural layering in complex organisations can undermine responsiveness by spreading accountability, slowing decision-making, and reducing operational coherence (Weick & Sutcliffe, 2007). Power (2007) similarly argues that modern risk-management systems can produce "organised uncertainty," in which increased oversight does not necessarily translate into better risk control. In AI governance, additional procedures may enhance visibility without improving the organisation's actual capacity to manage effectively.

This problem is reinforced by external technological dependency: reliance on proprietary AI systems, cloud services, and vendor-controlled models limits internal governance even as regulatory accountability remains internal (Burrell, 2016), so additional procedures aimed at compliance visibility leave the underlying constraints on technical access, intervention authority, and operational coordination unaddressed. Most existing literature assumes that stricter regulation and formal governance straightforwardly improve oversight; research on institutional complexity suggests otherwise, indicating that governance expansion can instead produce fragmentation, procedural overload, and reduced responsiveness under technological uncertainty.

## CONCEPTUAL METHOD AND THEORY-BUILDING APPROACH

The paper presents a theoretical framework based on institutional theory, research on organisational governance and accountability, and emerging literature on AI governance. Instead of empirically testing causal relationships, it explains how growing regulatory complexity can unintentionally undermine organisational control in environments that heavily rely on AI.

Conceptual research is crucial in rapidly evolving technological fields where institutional conditions change faster than stable empirical models can be developed (Jaakkola, 2020). AI governance exemplifies this scenario. While there has been considerable research on ethical principles, accountability frameworks, transparency mechanisms, and responsible AI practices (Floridi & Cowls, 2019; Morley et al., 2020; Wieringa, 2020), there has been less focus on how governance expansion affects organisational control capacity, especially in technologically complex and externally dependent contexts.

The study builds the Governance Inversion Hypothesis by integrating institutional theory, organisational complexity, socio-technical systems research, and AI governance literature, reinterpreting existing theories of institutional decoupling, procedural complexity, accountability fragmentation, and technological opacity in the context of AI-intensive governance. It follows a configurational theory-building approach, identifying the interconnected institutional dynamics detailed in the four mechanisms below through which governance formalisation can increase while operational control diminishes.

The paper does not assert that governance inversion happens in every organisation or regulatory setting. Instead, it offers a framework to guide future research across various sectors, institutional structures, and AI governance models. The propositions that follow are exploratory and aim to generate theory rather than establish predictive laws.

## THE GOVERNANCE INVERSION HYPOTHESIS (GIH)

### From Regulatory Expansion to Control Reduction

Conventional governance theory posits that more regulation enhances organisational control. Regulatory systems aim to reduce uncertainty, boost accountability, and improve oversight via formal governance structures, monitoring processes, reporting requirements, and compliance mechanisms (Bovens, 2007; OECD, 2022).

This assumption becomes less stable in AI-intensive environments. AI systems are distinct from traditional organisational technologies. They utilise adaptive learning, complex processes, distributed infrastructures, and dynamic external ecosystems (Burrell, 2016; Rahwan, 2018). As AI adoption increases, organisations must show responsible governance through ethics frameworks, reporting systems, audits, compliance measures, and oversight structures. Enhancing governance, nonetheless, does not always improve operational authority. We present the Governance Inversion Hypothesis. It suggests that as AI complexity and regulatory demands grow, broader governance structures unintentionally reduce effective control over AI systems.

The Governance Inversion Hypothesis (GIH) is articulated as follows:

> *Increasing regulatory pressure in AI-intensive environments may expand governance visibility while weakening effective organisational control over AI systems.*

The hypothesis posits that AI-intensive systems alter the dynamics of regulation and control. Specifically, it highlights that when the expansion of formal governance outstrips an organisation's ability to manage operations, governance may increase while operational authority becomes fragmented, outsourced, and restricted by procedures.

As regulatory pressures increase, organisations often add layers of governance, such as committees, ethics frameworks, and compliance functions. These added structures can weaken authority, slow down escalation processes, and increase reliance on external technology. This dependence intensifies when organisations turn to external AI systems. Hence, while accountability stays within the organisation, operational authority shifts to vendors and proprietary AI ecosystems.

There are two key distinctions of governance formalisation and governance control.

- Governance formalisation is the visible establishment of formal structures like AI committees, responsible AI frameworks, ethics policies, governance roles, and compliance systems.
- Governance control is the organisation's actual ability to exercise authority over AI system behaviour, deployment, modifications, risk exposure, and decision outcomes.

The GIH argues that these two dimensions increasingly diverge.

The framework consists of four interconnected mechanisms:

1. Fragmentation of authority,
2. Symbolic governance expansion,
3. Externalisation of control,
4. Authority paralysis.

***Mechanism I: Fragmentation of Authority***

As regulations on AI systems increase, organisations often spread governance roles across departments, including compliance, legal, risk management, ethics, cybersecurity, data governance, audit, and tech operations. While this broadens governance, it can also reduce operational coherence by scattering decision-making authority across overlapping layers.

Traditional governance models rely on clear authority structures to ensure accountability (Bovens, 2007), but in AI-intensive settings, governance responsibilities overlap across technical, legal, operational, and regulatory domains. For instance, an AI-driven lending system involves technology teams handling infrastructure, compliance units interpreting regulations, legal departments managing liability, data governance teams overseeing datasets, and operational managers implementing results. As a result, authority is distributed rather than concentrated.

Research indicates that institutional layering in complex organisations can weaken decision-making and blur accountability, particularly in bureaucratic systems (Crozier, 1964; Simon, 1972). This issue is exacerbated in AI governance, where organisations often add oversight structures in response to regulations rather than rethinking authority relationships. This can lead to overlapping jurisdictions, fragmented escalation pathways, and unclear operational ownership.

Studies on algorithmic accountability confirm the challenges of assigning responsibility in complex socio-technical systems (Burrell, 2016; Wieringa, 2020). Raji et al. (2020) state that effective AI accountability relies on coherent institutional authority, yet many organisations disperse AI governance duties across specialised functions without centralised operational control. The problem is particularly evident in highly regulated sectors like banking, healthcare, insurance, and public administration. In these areas, the addition of AI governance complicates existing compliance and oversight systems (Frimpong & Botchey, 2026). This expansion can hinder responsiveness by disrupting coordination, slowing escalation processes, and spreading intervention authority too thin.

The Governance Inversion Hypothesis proposes:

**Proposition 1 (P1):** *Increasing regulatory pressure is positively associated with the expansion of formal AI governance structures but negatively associated with the concentration of operational decision authority over AI systems.*

***Mechanism II: Symbolic Governance Expansion***

Organisations are increasingly establishing visible AI governance structures to demonstrate responsibility and regulatory compliance. These structures include ethics committees, frameworks for responsible AI, governance roles, audit procedures, and formal reporting systems.

According to institutional theory, organisations often adopt formal governance arrangements in response to external pressures, even when these structures are not fully integrated into their operations (Meyer & Rowan, 1977). In the realm of AI governance, this is compounded by rising expectations from regulators, investors, clients, and the public concerning responsible oversight (OECD, 2022; Papagiannidis et al., 2025).

However, governance formalisation does not always confer operational authority. Symbolic governance structures enhance institutional legitimacy but remain disconnected from key decisions and processes, such as deployment, technical interventions, or operational escalations. Examples include advisory committees lacking veto power, ethical principles without enforcement, and governance roles separate from core operational decisions.

Research on AI governance has shown significant gaps between ethical principles and their practical application. Mittelstadt (2019) notes that ethical principles alone cannot ensure responsible AI outcomes. Morley et al. (2020) and Prem (2023) emphasise the limited integration of many governance tools into decision-making processes.

The GIH posits that heightened regulatory pressure leads organisations to focus more on procedural visibility than on enhancing their operational capabilities. As a result, governance systems become more visible at an institutional level while still facing operational limitations.

The GIH presents the following proposition:

**Proposition 2 (P2):** *Growing regulatory pressure is associated with the expansion of symbolic AI governance structures that increase institutional visibility without proportionately strengthening operational control.*

***Mechanism III: Externalisation of Control***

Organisations are becoming more reliant on externally sourced AI infrastructures. While they maintain formal accountability, they often lose direct control over these systems. As a result, operational authority shifts to vendors, cloud providers, and proprietary infrastructure, leaving organisations unable to fully build or understand the AI systems they use.

The trend is evident in regulated sectors, where organisations use externally sourced AI for tasks such as credit scoring, fraud detection, underwriting, diagnostics, cybersecurity, and automated decision support. Although these organisations are responsible for AI outcomes, they often have limited insight into the models, training data, optimisation processes, or system architectures involved.

Research on algorithmic opacity reveals that machine-learning systems are often difficult to interpret and govern effectively (Burrell, 2016). The *Black Box Society* similarly argues that contemporary digital systems increasingly operate through opaque institutional and technological infrastructures that limit transparency, accountability, and meaningful external oversight (Pasquale, 2015). Rahwan (2018) further highlights that autonomous socio-technical systems challenge traditional assumptions surrounding organisational supervision and control.

This governance assumption becomes increasingly unstable under conditions of external technological dependency. Vendor-controlled AI systems continuously evolve through updates and optimisations that organisations may not fully observe or control. As a result, governance relies heavily on external tech actors beyond the organisation's direct oversight.

There is a structural imbalance between accountability and authority in organisations concerning AI outcomes. While organisations are responsible for these outcomes, control is often spread across external technologies. The GIH posits that as regulations grow more complex, organisations increasingly depend on external compliance tools, vendor governance systems, cloud platforms, and third-party AI infrastructures to meet these governance demands.

Governance visibility consequently expands institutionally while operational authority becomes progressively externalised.

The GIH posits the following proposition:

**Proposition 3 (P3):** *Increasing reliance on external AI infrastructures weakens the alignment between organisational accountability and operational control in highly regulated sectors.*

***Mechanism IV: Authority Paralysis***

As AI governance systems become more complex, organisations find it challenging to keep pace with rapidly changing technologies. AI systems adapt to and process real-time data, while governance often relies on slower processes such as committee reviews and compliance checks. Research shows that complex procedures can hinder responsiveness in high-pressure situations (Perrow, 1984; Weick & Sutcliffe, 2007). In environments that rely heavily on AI, increased governance slows urgent response efforts precisely when quick coordination is needed. In highly regulated sectors, adding AI governance on top of existing compliance and supervisory structures creates significant problems. Organisations may experience delays in decision-making, unclear escalation responsibilities, and fragmented authority for interventions.

Authority paralysis occurs when governance actors have oversight roles but lack clear authority or decision-making power. Fears of liability, regulatory uncertainty, and overlapping responsibilities can lead to delays in taking action. As governance systems become more procedural, their ability to respond quickly may diminish.

AI governance research points to similar issues. Shneiderman (2020) asserts that trustworthy AI requires meaningful human control amid technical complexity, while Koshiyama et al. (2024) emphasise the institutional capabilities needed for effective oversight and intervention.

The GIH posits the following proposition:

**Proposition 4 (P4):** *Increasing governance complexity and procedural layering reduce the speed and coherence of organisational intervention in AI-intensive systems.*

***Integrating the Mechanisms: The Inversion Pathway***

These four mechanisms show how regulatory pressure, in increasing accountability and oversight structures, can simultaneously fragment authority, promote symbolic compliance, shift operational control externally, and diminish the timeliness of intervention. This pattern is consistent with evidence that AI governance in many organisations remains advisory, weakly institutionalised, and poorly positioned within executive decision-making (Frimpong & Botchey, 2026). Figure 1 traces this inversion pathway, showing how formalising governance intended to strengthen accountability and oversight, instead trigger interconnected mechanisms that reduce operational coherence and intervention capability in AI-intensive settings.

**Figure 1**. The Governance Inversion Pathway
The central logic of the Governance Inversion Hypothesis

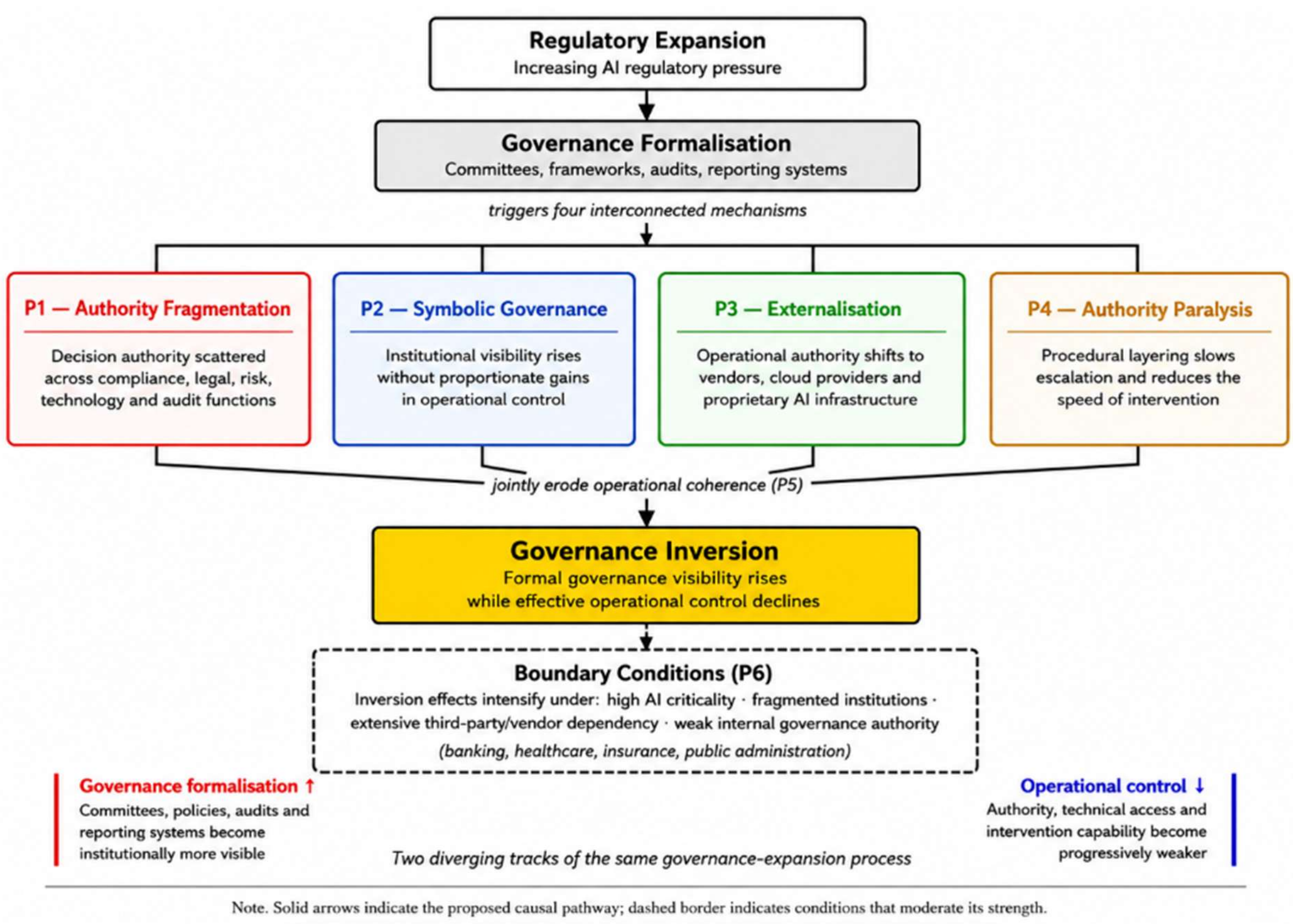

Source: The Author, 2026

Figure 1 visualises the core logic of the Governance Inversion Hypothesis as a single pathway rather than a static hierarchy. Regulatory expansion is shown driving governance formalisation, which in turn activates the four mechanisms specified in P1–P4; authority fragmentation, symbolic governance expansion, externalisation of control, and authority paralysis, each rendered as a distinct panel to emphasise that these are parallel, co-occurring processes rather than a single sequential cause. The convergence of all four pathways into the governance inversion outcome reflects P5: that inversion effects intensify where formal governance expansion is not matched by a corresponding concentration of operational authority.

The figure further distinguishes two diverging tracks running beneath the same regulatory-expansion process: governance formalisation, which becomes institutionally more visible through committees, policies, and reporting systems, and operational control, which becomes progressively weaker as authority, technical access, and intervention capability erode.

Representing these as separate, opposing tracks, rather than a single degradation curve, makes explicit the structural claim at the centre of the GIH: that the two dimensions are not simply uncorrelated but move in opposite directions under the conditions the paper specifies. The dashed boundary-conditions panel corresponds to P6, indicating that the strength of the inversion effect is not uniform but is moderated by AI criticality, institutional fragmentation, third-party dependency, and the weakness of pre-existing governance authority, conditions most pronounced in banking, healthcare, insurance, and public administration.

The framework proposes:

**Proposition 5 (P5):** *Organisations exhibiting high levels of formal governance expansion without corresponding concentration of operational authority are more likely to experience governance inversion effects.*

## Boundary Conditions and Scope of the Hypothesis

The GIH is most applicable when there is high AI complexity, fragmented institutions, external reliance on technology, and weak governance authority.

### *AI criticality*

Governance inversion is more likely in high-stakes environments like banking, healthcare, insurance, recruitment, law enforcement, and public administration. These sectors face significant regulatory pressure while relying on complex, hard-to-interpret AI systems (Burrell, 2016; Koshiyama et al., 2024).

### *Organisational complexity*

Large institutions typically rely on complex governance structures that include legal, compliance, cybersecurity, audit, ethics, procurement, and technology functions. However, in AI-intensive environments, these structures can dilute operational authority rather than enhance it. Recent findings indicate that many AI governance roles are primarily advisory, lack adequate resources, and are not well established (Frimpong & Botchey, 2026).

### *External technological dependency*

Governance issues escalate when organisations rely heavily on proprietary AI systems, cloud services, vendor APIs, or externally managed foundation models. While accountability stays within the organisation, operational control shifts to external technology.

### *Symbolic compliance pressure*

Organisations facing significant reputational or regulatory pressure often focus on governance formalisation through policies, committees, reporting systems, and ethics programs. These measures are aimed more at demonstrating legitimacy than enhancing operational control. This aligns with institutional theory, which posits that organisations create formal structures for legitimacy without fully incorporating them into their daily operations (Meyer & Rowan, 1977).

***Weak veto authority***

Governance actors might oversee AI systems but often lack the authority to halt deployments, contest outputs, demand technical changes, or prevent decisions from being implemented. Consequently, their role is more consultative than authoritative.

The framework does not argue against regulation; rather, it cautions against the assumption that regulation itself ensures control.

The paper proposes:

**Proposition 6 (P6):** *The relationship between regulatory expansion and declining organisational control is strengthened under conditions of high AI complexity, strong third-party dependency, and weak internal governance authority.*

Table 1 outlines the propositions from the GIH and connects each proposition to its theoretical basis.

**Table 1**

Summary of Propositions Developed from the Governance Inversion Hypothesis

| Proposition | Mechanism | Core Claim | Observable Indicator |
|---|---|---|---|
| P1 | Authority fragmentation | Regulatory pressure expands formal AI governance structures while reducing concentration of operational decision authority. | Multiple governance committees, overlapping reporting lines, unclear veto authority, duplicated oversight functions, fragmented escalation pathways. |
| P2 | Symbolic governance expansion | Regulatory pressure increases the visibility of governance structures without a proportionate improvement in operational control. | Ethics frameworks without enforcement power; advisory governance roles; compliance-oriented reporting systems; governance policies disconnected from operational intervention. |
| P3 | Externalisation of control | Reliance on external AI infrastructures weakens alignment between organisational accountability and operational control. | Dependence on proprietary vendors, cloud-based AI services, limited access to model architecture or training data, restricted auditability of third-party systems. |
| P4 | Authority paralysis | Procedural layering reduces the speed and coherence of organisational intervention in AI-intensive systems. | Slow escalation processes, delayed intervention decisions, excessive layers of approval, and governance bottlenecks during AI incidents or system changes. |
| P5 | Governance inversion effect | Organisations with high governance formalisation but weak operational authority are more likely to experience governance inversion. | High governance formalisation combined with weak intervention capability, low authority concentration, fragmented accountability, and an inability to effectively override AI decisions. |
| P6 | Boundary conditions | AI complexity, third-party dependency, and weak internal authority strengthen the regulation–control inversion effect. | High-risk AI deployment, extensive vendor dependency, opaque AI systems, weak institutionalisation of governance, limited technical oversight capacity. |

Table 1 presents the GIH as a framework that emphasises interconnected institutional mechanisms rather than a straightforward causal relationship. It explains how increasing regulatory measures can unintentionally weaken organisational control. The framework includes observable indicators that link each proposition to measurable organisational conditions, enhancing its empirical relevance. Rather than viewing governance as merely formal structures, it focuses on the relationship between governance formalisation and operational authority. This distinction will allow future research to examine whether organisations that appear well-governed, in fact, face declining intervention capability, fragmented authority, and reduced control over AI systems.

## THEORETICAL IMPLICATIONS

### Reframing AI Governance as a Problem of Control

Current research on AI governance emphasises key areas such as ethics, accountability, transparency, fairness, explainability, and compliance (Floridi & Cowls, 2019; Mittelstadt, 2019). This has led to the creation of governance principles, regulatory frameworks, and guidelines for responsible AI. However, it often assumes that organisations have the necessary capacity to implement these governance measures effectively.

The GIH questions this assumption by shifting focus from procedural compliance to organisational control capacity. In AI-intensive environments, formal governance structures may grow, but operational authority can be weakened. Organisations can establish ethics committees, audit systems, governance roles, reporting procedures, and compliance frameworks, but these alone do not guarantee proper authority over the operational management of AI systems. The framework conceptualises AI governance as a problem of aligning control across accountability, authority, technical access, and intervention capability.

This approach builds on current AI governance literature in significant ways.

- It challenges the assumption that governance expansion necessarily strengthens organisational oversight. Existing governance frameworks frequently imply a linear relationship between regulation and control (OECD, 2022; NIST, 2023). The GIH instead proposes that the expansion of governance eventually weakens supervisory coherence under conditions of institutional complexity.
- The framework emphasises governance effectiveness over mere existence. While many organisations have established visible AI governance structures, evidence indicates that these frameworks are often poorly institutionalised, advisory, and disconnected from operational authority (Frimpong & Botchey, 2026).
- The framework views AI governance as an organisational design issue rather than just a compliance matter. Its effectiveness relies on how authority, escalation pathways, technical visibility, and intervention rights are allocated within institutions.

This change significantly affects highly regulated industries like banking, healthcare, insurance, and public administration. These sectors are adopting complex governance structures while depending on opaque, externally sourced AI systems. This governance complexity could lead to a loss of control.

The GIH redefines governance failure as a structural issue of institutional control instead of merely a failure of ethics or regulatory compliance.

**Extending Institutional Theory: From Decoupling to Inversion**

Institutional theory explains how organisations adopt formal governance structures to maintain legitimacy under external pressure (Meyer & Rowan, 1977). These structures often signal conformity to regulatory and societal expectations even when loosely connected to operational practice.

The GIH extends this logic into AI-intensive environments. Traditional decoupling theory suggests that formal governance structures and operational practices become disconnected. The GIH goes further, arguing that the expansion of governance itself weakens operational control amid institutional complexity.

This dynamic is increasingly evident in AI governance systems that operate through layered oversight structures encompassing compliance, legal, ethics, cybersecurity, audit, procurement, data governance, and executive management. As governance architectures expand, operational authority becomes increasingly dispersed across institutional layers.

The problem is especially acute in highly regulated sectors. Organisations frequently respond to escalating governance expectations by adding new governance structures to existing institutional systems rather than fundamentally redesigning authority relationships. The result is growing institutional density without corresponding integration of operational control.
Organisational studies have similarly shown that institutions preserve formal legitimacy structures while operational vulnerabilities accumulate internally (Vaughan, 1997). In AI governance environments, expanding procedural oversight therefore coexists with weakening supervisory coherence.

The framework introduces inversion rather than simple decoupling. Governance structures do not merely become disconnected from operational practice. Under inversion conditions, governance expansion actively contributes to the erosion of control by dispersing authority, slowing intervention pathways, and increasing procedural complexity.

The GIH also departs from Power’s (2007) account of organised uncertainty, with which it shares a scepticism toward procedural risk management but not its underlying mechanism. Power’s argument is epistemic: proliferating audit and risk-management procedures multiply the categories through which risk is defined and documented, without necessarily reducing the underlying uncertainty they are meant to manage, so that organisations become oriented toward auditability and defensibility rather than toward risk itself. The GIH identifies a distinct and more specifically structural mechanism. Its claim is not that procedures fail to resolve uncertainty, but that governance expansion actively relocates operational authority away from the organisation, dispersing it across internal functions (P1), diverting it into symbolic structures (P2), transferring it to external vendors (P3), and slowing its exercise through procedural layering (P4), such that formal oversight and the capacity to act increasingly diverge. Organised uncertainty describes a condition of the knowledge governance produces;

governance inversion describes a condition of the authority governance redistributes. The two dynamics may co-occur an organisation that manages risk through proliferating categories is also more likely to disperse the authority needed to act on them but they are conceptually separable, and the GIH treats authority relocation, rather than risk categorisation, as the operative mechanism linking regulatory expansion to declining operational control.

The GIH also introduces the concept of control alignment. Governance effectiveness depends on maintaining coherent relationships between accountability, authority, technical access, and intervention capability. AI governance should therefore be evaluated not only by institutional visibility, but by whether governance systems preserve meaningful organisational control under conditions of technological complexity and external dependency.

This extension contributes to institutional theory by reframing legitimacy-oriented governance structures as potential sources of operational weakening in AI-intensive environments.

**Rethinking Accountability in AI Governance**

AI governance research frequently treats accountability as a problem of transparency, explainability, documentation, or human oversight (Wieringa, 2020; Raji et al., 2020). Existing frameworks generally assume that organisations can maintain accountability if sufficient governance procedures are established around AI systems.

The GIH challenges this assumption at its core. Accountability failures in AI systems may arise from structural misalignment rather than from informational opacity alone. This misalignment arises when organisations remain formally responsible for AI outcomes while lacking the operational authority to govern the systems that generate them. Accountability obligations remain organisationally internal even as operational control becomes fragmented, externalised, or procedurally constrained.

The framework points out the limitations of current AI auditing models. While governance frameworks are increasingly focusing on algorithmic auditing and accountability (Koshiyama et al., 2024), effective auditing requires strong authority, technical access, operational independence, and the ability to intervene. Currently, dedicated AI auditing functions are poorly established, even in well-regulated industries.

The GIH shifts the focus of accountability from procedural oversight alone to an organisation's ability to control itself. Effective accountability systems require cohesive authority relationships, clear operational visibility, and strong intervention capabilities. This perspective clarifies why visible accountability structures can exist alongside governance failures. Organisations meet procedural accountability requirements but still lack the ability to manage AI systems effectively. This framework shifts accountability discussions from mere disclosure to more critical issues of organisational authority, governance structure, and operational control.

### Regulation and Governance Design

Current AI regulation largely adopts a visibility-centred model of governance that prioritises formal procedures such as policies, audit systems, ethics frameworks, documentation requirements, reporting mechanisms, and governance committees. While these mechanisms remain important for accountability and institutional legitimacy, their expansion does not necessarily strengthen operational control. Organisations can often demonstrate procedural compliance more easily than coherent authority, technical oversight, or effective intervention capability.

This creates a structural governance risk in AI-intensive environments. As governance requirements expand, institutions prioritise procedural conformity over developing operational control capacity. In highly regulated sectors, additional governance layers increase institutional complexity without improving responsiveness, escalation capability, or supervisory coherence. The problem becomes more pronounced under conditions of external AI dependency, where accountability remains internal to the organisation while operational authority is increasingly distributed among cloud providers, proprietary model developers, infrastructure vendors, and external AI ecosystems.

The GIH challenges the assumption that greater governance automatically yields greater control. Under conditions of technological complexity, procedural expansion contributes to governance saturation, fragmented authority, and slower institutional intervention. Effective AI governance consequently depends not only on formal governance structures, but on whether organisations retain concentrated authority, technical access, audit independence, and meaningful intervention capability over increasingly autonomous socio-technical systems.

## MANAGERIAL AND POLICY IMPLICATIONS

### Managerial Implications: Governing for Control

The GIH argues that organisations should focus on operational control rather than simply expanding governance structures. While many have set up AI committees and ethics frameworks, governance actors often lack direct authority over deployment decisions and technical interventions (Frimpong & Botchey, 2026). For effective AI governance, it is essential that these functions have real decision-making power, not just advisory roles.

Managers need to enhance governance by clearly defining operational ownership. AI governance structures must include clear escalation pathways, concentrated veto authority, and direct access to technical information. Distributing governance responsibilities across various compliance, legal, audit, and operational functions can slow down responses during AI incidents or model failures.

The framework emphasises the growing need for audit capabilities in AI environments that rely on external models and vendors. Organisations depend on proprietary algorithms and cloud systems, yet remain accountable for their results. As a result, managers should seek stronger contractual audit rights, access to models, vendor transparency, and independent validation to ensure proper oversight of these AI systems.

Moreover, governance systems must focus on intervention capabilities. Effective oversight requires the authority to pause deployments, challenge model outputs, address concerns quickly, and demand technical fixes when necessary. AI governance is less effective when governance bodies can identify risks but lack the ability to act on them.

More broadly, AI governance should be seen as a strategic control function within organisations, not just a compliance task. Its effectiveness relies on strong authority, effective coordination, clear technical understanding, and the ability to act in complex technological situations.

**Policy Implications: The Limits of Procedural Governance**

The GIH has key implications for the design of AI regulation. Current frameworks focus heavily on procedural accountability, like documentation, reporting, ethics guidelines, audits, and transparency. While these are important, just expanding procedures doesn't ensure effective organisational control.

Future AI regulation should prioritise governance capability over formal governance. Regulatory assessments need to evaluate whether organisations have concentrated operational authority, the ability to escalate issues, technical access, independent audits, and effective rights of intervention over AI systems.

The framework emphasises the growing governance risks associated with reliance on external technology. Many organisations use AI systems that rely on third-party infrastructure, proprietary models, and external cloud services, yet they remain accountable for the outcomes. Regulatory systems need to implement stricter requirements for vendor transparency, audit access, explainability, and the right to challenge or suspend these external AI systems.

The findings indicate that excessive procedural layers could hinder governance responsiveness, especially in highly regulated sectors. Future regulations should aim to balance accountability with operational flexibility, particularly in situations that demand quick action and coordinated oversight.

The paper highlights that effective AI governance depends not only on the existence of governance frameworks, but on whether institutions retain meaningful authority over increasingly complex and externally mediated AI infrastructures.

## FUTURE RESEARCH DIRECTIONS

The GIH serves as a framework for building theories rather than a definitive explanatory model. The paper highlights key areas for future research in organisational control, institutional design, and AI governance amidst technological complexity.

The hypothesis needs systematic empirical testing across various sectors, organisational structures, and regulatory environments. Evidence indicates fragmented governance, weak oversight, and inconsistent institutionalisation of AI regulation (Frimpong & Botchey, 2026). Future research should investigate whether greater regulatory intensity leads to reduced operational control over AI systems. Comparative studies across banking, healthcare, insurance, public administration, and technology would be beneficial given their distinct governance structures and risk profiles.

Future research should focus on empirically measuring governance inversion. The framework proposed here identifies key dimensions, including governance formalisation, authority concentration, intervention capability, vendor dependency, governance fragmentation, and operational responsiveness. These dimensions could help create a Governance Inversion Index to evaluate the gap between formal governance expansion and actual organisational control across institutions.

Future research should explore the broader implications of governance inversion. Current AI governance studies mainly focus on ethical principles, transparency, and accountability (Papagiannidis et al., 2025; Prem, 2023). In contrast, the GIH indicates that governance effectiveness depends on an organisation's capacity for control. Future studies should examine how governance inversion affects organisational resilience, institutional legitimacy, operational failures, audit capabilities, and accountability systems in increasingly autonomous socio-technical environments.

The future of AI governance research should shift from procedural analysis to a deeper examination of authority alignment, organisational control, and institutional capacity, especially as technological complexity increases.

## CONCLUSION

This paper has introduced the Governance Inversion Hypothesis to explain a structural tension increasingly visible in AI governance: organisations can become more formally governed while simultaneously losing operational control over the AI systems they are meant to oversee.

This challenges the common assumption that expanding governance frameworks automatically improves organisational oversight. Current AI governance research and regulatory models often treat more formal governance as synonymous with better control; the GIH suggests this relationship can break down in AI-heavy environments marked by institutional complexity, fragmented authority, external technology dependence, and opaque socio-technical systems. The framework reframes AI governance from a procedural or ethical exercise into a problem of organisational control alignment — one requiring not just policies and committees, but coherent authority, technical visibility, escalation capability, audit access, and the power to intervene meaningfully.

The paper makes three main contributions: it introduces governance inversion as a concept explaining how governance expansion can unintentionally weaken organisational control in complex technological environments; it extends institutional theory beyond decoupling by showing how formal governance can itself produce operational fragmentation; and it reframes AI accountability failures as structural problems of misalignment between responsibility and control, a misalignment sharpened by growing reliance on proprietary, vendor-controlled AI infrastructure that leaves organisations accountable for systems they do not fully control.

Future AI governance frameworks, the GIH suggests, should prioritise operational coherence and concentrated authority alongside procedural transparency. The central challenge is not the absence of governance structures, but sustaining the capacity to govern effectively as AI systems become more autonomous and externally mediated.